\documentclass[12pt]{amsart}
\usepackage[margin = 1in]{geometry}
\usepackage{subfig, graphicx, multirow, multicol}
\usepackage{amsmath, amssymb, bm, booktabs}
\usepackage{hyperref}
\usepackage{algorithm}
\usepackage{algpseudocode}
\nocite{*}
\allowdisplaybreaks

\begin{document}
\title[Three-Stage Bubble Analysis]{A three-step machine learning approach to predict market bubbles with financial news}
\author{Abraham Atsiwo}
\address{Department of Mathematics \& Statistics; University of Nevada, Reno}
\email{aatsiwo@unr.edu}

\begin{abstract}
This study presents a three-step machine learning framework to predict bubbles in the S\&P 500 stock market by combining financial news sentiment with macroeconomic indicators. Building on traditional econometric approaches, the proposed approach predicts bubble formation by integrating textual and quantitative data sources. In the first step, bubble periods in the S\&P 500 index are identified using a right-tailed unit root test, a widely recognized real-time bubble detection method. The second step extracts sentiment features from large-scale financial news articles using natural language processing (NLP) techniques, which capture investor’s expectations and behavioral patterns. In the final step, ensemble learning methods are applied to predict bubble occurrences based on both sentiment-based and macroeconomic predictors. Model performance is evaluated through k-fold cross-validation and compared against benchmark machine learning algorithms. Empirical results indicate that the proposed three-step ensemble approach significantly improves predictive accuracy and robustness, providing valuable early warning insights for investors, regulators, and policymakers in mitigating systemic financial risks.
\end{abstract}

\maketitle

\newtheorem{theorem}{Theorem}
\newtheorem{lemma}{Lemma}
\theoremstyle{definition}
\newtheorem{asmp}{Assumption}
\newtheorem{defn}{Definition}
\newtheorem{example}{Example}
\thispagestyle{empty}

\section{Introduction}

%\subsection*{Bubbles and Market Crashes}
The aftermath of market bubbles and financial collapses is of paramount interest to stakeholders due to the damage they cause to the mindset of the average investor. Investors are cautious after recovering from an economic recession due to the instability they cause.
The United States alone has experienced several financial collapses and market bubbles in the last 100 years: the 1929 Wall Street crash, the 1937 recession, the 1987 Black Monday crash, the 1990 early recession, the financial crisis of 2007-2008, the 2015-2016 stock market sell-off, and the 2020 coronavirus market crash \cite{mccarty2013political}. 
In addition to the USA, approximately 100 crises have occurred in the last 30 years worldwide \cite{akerlof2014have}.

Institutions and researchers have explored non-technical approaches to explaining and examining market bubbles and financial collapses. Market bubbles and financial crises can be attributed to different factors. A "political bubble" stems from an unstable political environment, which leads to speculative behavior and further drives the market. Some of the above-mentioned crashes can be predicted by examining the prices of financial derivatives \cite{bates1996dollar}, \cite{johansen2010shocks}. However, we will consider market bubbles and the financial crisis from a mathematical point of view, that is, observing the history of prices to detect the occurrence of bubble instances. This path contradicts the semi-strong form of the efficient market hypotheses which assumes that stock price movements are a reflection of publicly available information.  In general, the pricing of assets and financial derivatives can be modeled by some complex mathematical equations. The Black-Scholes model is used to determine the fair price of European-style options and derivatives, the Binomial Option Pricing model is useful for pricing American options, and the Monte Carlo simulation is used to approximate complex mathematical functions. The price of financial instruments does not always follow their fundamental pricing equation, leading to inequity in financial markets. A bubble is said to occur if the price of a financial derivative deviates from its fundamental according to the pricing model \cite{sornette2009stock}. 

%Most bubbles and financial collapses can be attributed to a deviation from market fundamentals, and thus mathematical approaches cannot be neglicted. 

The main goal of this study is to predict market bubbles with economic indicators and financial news in a multilabel classification approach using an ensemble method. The S\&P 500, a proxy for the US stock market, has global implications and will therefore be used in this analysis. The three-step approach also selects important variables that determine the occurrence of a market bubble and financial collapse, providing an early warning signals for all stakeholders.

\subsection*{Bubble Detection}
The Augmented Dickey-Fuller (ADF) test\cite{dickey1979distribution} was proposed  to test the presence of a unit root in autoregressive time series models. The autoregressive parameter $\rho$ estimated with regression techniques was their main parameter of interest. If $\rho = 1$, then the time series is non-stationary. On the other hand, if $|\rho| > 1,$ then the given series has a variance that grows exponentially in addition to being non-stationary. Dickey and Fuller provide an analytical form of the asymptotic distribution, which serves as a useful tool for testing hypotheses and constructing confidence intervals in econometric applications. Despite its widespread use, it cannot detect whether a given observation represents a bubble period or not. \cite{chang2002asymptotics} derived the asymptotic distribution of the ADF test under weaker conditions and \cite{galbraith1999distributions} considered the distribution of the ADF test in the presence of moving average components. The ADF test and its asymptotic distribution serve as a natural starting point for bubble detection despite its limitations. 

\cite{phillips2011explosive}, \cite{phillips2015testinga}, \cite{phillips2015testing} developed recursive approaches to bubble detection for econometric applications. The PWY procedure uses a forward recursive sliding window approach to capture periodic collapsing bubbles of \cite{evans1991pitfalls}. Unlike the ADF test, the PSY, also known as the supremum augmented Dickey-Fuller test (SADF), is a right-tailed test where the alternative hypothesis suggests the presence of an explosive root. The PWY procedure cannot address the economic challenge associated with long sequences. Long sequences tend to be non-linear and exhibit structural breaks in their occurrence. The main drawback of the PWY procedure that inhibits its use is its inability to detect multiple bubbles. 

The Philip Shi and Yu (PSY) approach, like PWY, is a recursive approach that uses a sliding-window approach to detect the occurrence of market bubbles and financial collapses in real time. The PSY approach uses a forward and backward recursive technique to detect the beginning and end of bubbles and market crashes. Unlike the PWY procedure, the PSY procedure is capable of detecting multiple bubbles associated with long sequences. It is able to detect non-linear patterns and structural breaks known with long sequences. This approach was able to detect famous market bubbles and financial crashes when applied to the S\&P 500 market index. It is the benchmark for determining the origination and termination of market bubbles and collapses in the econometric literature. It serves as a warning alert for surveillance teams and central banks.

\subsection*{Financial News for Stock Price Movement}
Textual data for predicting stock prices come in various forms: financial and general news, social networks, company announcements, and blog sites. We review literature on predicting stock prices using three main data sources: data from corporate announcements, financial and general news and social media, and microblogging. 

The rapid rise and use of social networks and blogging websites have created an enormous volume of data. These sources of data can be accessed through an API or by web-scrapping if allowed. Users and influential people share financial and political news feeds on these platforms. These data are important in predicting stock movement in the short term \cite{yuan2020dancing}. Social media data are used mainly in studies related to sentiment analysis. \cite{bollen2011twitter} examine the correlation between stock price movement and public sentiments with Twitter data, which aligns with the assertion of behavioral economists that emotions can affect individual decision making. \cite{smailovic2014stream} and \cite{li2017discovering} used Twitter data to predict stock price movement in "Stream-based active learning for sentiment analysis in the financial domain" and "Discovering public sentiment in social media to predict stock movement of publicly listed companies".

\cite{shi2018deepclue} in "DeepClue: Visual Interpretation of Text-Based Deep Stock Prediction" presents an approach to bridge the gap between text-based deep learning techniques and end users using financial news from Reuters and Bloombery. Their approach provides a visualization interface and interprets key factors learned by the hierarchical deep learning model. \cite{peng2015leverage} leveraged word embeddings in the Reuters and Bloombery news feeds and a deep neural network to predict the movement of stock prices in the market. The proposed approach outperforms models that uses historical price information. \cite{schumaker2012evaluating} used data from Arizona Financial Text (AZFinText) system to develop a sentiment analysis tool to measure the impact of financial news on investors sentiment. Wu et al . \cite{wu2012stock} use both technical and news data for prediction in a more traditional approach, Linear regression.

Textual data from corporate disclosure are more accurate compared to news from social media and news feeds from third-party vendors. News from corporate disclosure include quarterly earnings, layoffs, semi-annually and annually financial statements, and legal disclosure.  \cite{feuerriegel2018long} utilizes predictive models for high-dimensional data to predict the long-term (24 months) impact of regulatory disclosure on stock indices. \cite{groth2011intraday} employ a strategy for managing market risk intraday utilizing textual information. 

A summary of other articles using social networks, corporate earnings, and financial news is summarized in the review paper\cite{fataliyev2021stock}. 

\subsection*{Machine Learning and Deeping Learning Approaches}
The use of machine learning and deep learning approaches in predicting bubbles is rare in the econometric literature. The ML/DL based methods follow the laid-down structure: i) detect true labels from the training data, ii) use econometric variables as features, and iii) use an ML/DL approach to predict the true labels. 

\cite{bacsouglu2021two} uses a two-step approach to predict S\&P 500 bubbles. In their approach, the true labels were detected with the PSY procedure, economic indicators (GDP, BOP, Short-term Interest Rate, Long-Term Interest Rate and CPI) are used as predictors and an SVM is fitted to predict the true labels. \cite{ozgur2021detecting} detects speculative bubbles in metal prices with GSADF test and machine learning approaches. The Random Forest classifier shows that monetary rate and production index are important variables in predicting bubbles in metal prices. \cite{park2021machine} detects early warning signals for the housing and the stock market using LSTM. Their LSTM based approach detects changes in stock market volatility and house prices and performs better than random forest and SVM. \cite{ayan2021detection} detects and predicts price bubbles in Instanbul housing market using LSTM autoencoders. \cite{chatzis2018forecasting} forecast stock market crisis events using boosting and deep learning approaches. They employ boostrap sampling to handle imbalanced data associated with the true label. 

Despite extensive research on bubble prediction in the econometric literature, there are still gaps 
that need to be addressed. Studies \cite{bacsouglu2021two},  \cite{ozgur2021detecting}, \cite{chatzis2018forecasting}, \cite{ayan2021detection} and \\
\cite{park2021machine} do not distinguish between market crisis and market bubbles. Market bubbles occur when the price of an asset exceeds its fundamental value.  Asset prices see a steep decline in value when a financial crisis occurs. The financial crisis causes panic, which can lead investors to sell off their assets. None of the studies has examined the impact of financial news on bubble prediction. Building on the insights gained from this review, future research should prioritize the simultaneous impact of financial news and economic variables on bubble prediction.

\section{Methodology: The Three-Stage Analysis}

The three-stages of bubble analysis is discussed in this section: The PSY procedure for bubble detection, sentiment analysis with finbert-lc \cite{atsiwo2024financial} and also discussed in chapter two of \cite{atsiwo2024instruction}, and ensemble method for robust prediction. 

\subsection{ The PSY Procedure for Bubble Detection}
The PSY procedure \cite{phillips2015testing}, \cite{phillips2015testinga} is a recursive procedure used for the real-time identification and dating of financial bubbles. It is able to detect multiple bubbles in long financial series. The PSY procedure uses the generalized (GSADF) version of the SADF test, which was introduced in \cite{phillips2011explosive}. Like the SADF test, the GSDAF test uses a recursive approach with null and alternative hypotheses $$H_0: y_t = dT^{-\eta} + \theta y_{t-1} + \varepsilon_t ~~\text{vs}~~
H_1: \Delta y_t = \hat{\alpha}_{r_1, r_2} + \hat{\beta}_{r_1, r_2} y_{t-1} + \sum_{i=1}^k \hat{\phi}_{r_1, r2}^i \Delta y_{t-i} + \hat{\varepsilon}_t$$
% with test statistics
% \begin{align}
	%     % GSADF(r_0) = \sup_{r_2 \in [r_0, 1], r_1 \in [0, r_2-r_0]} ADF_{r_1}^{r_2}
	%      GSADF(r_0) = \sup_{\substack{r_2 \in [r_0, 1] \\ r_1 \in [0, r_2-r_0]} } ADF_{r_1}^{r_2}
	% \end{align}

The PSY procedure differs from the PWY procedure in the choice of starting point. The PSY procedure allows the start value $r_1$ of the recursion to be flexible. The start point of the GSADF test ranges between 0 and $r_2 - r_0$ and the test statistics is the supremum of the GSADF's test statistics
\begin{align}
	GSADF(r_0) =  \sup_{\substack{r_2 \in [r_0, 1] \\ r_1 \in [0, r_2-r_0]} } \left \{ ADF_{r_1}^{r_2} \right \}
\end{align}

At any point in time, the window size is $r_w = r_2 - r_1.$ Initially $r_1$ start from 0 and $r_2$ is allowed to range from $r_0$ to 1. At any time in the recursion $r_1$ varies from 0 to $r_2 - r_0$ and $r_2$ ranges from $r_0$ to 1. The moving window $[r_1, r_2]$ of length $r_w$ i used to estimate the $BSADF_{r_1}^{r_2}$ statistics at any point in time. The flexible start $r_1$ and end points $r_2$ allows this approach to detect the occurrence of multiple financial bubbles or market collapse.

First, we discuss binary classification for bubble prediction and then transition to multi-label classification. With binary classification for bubble prediction, true labels are identified from the given time series with a bubble detection method. The two classes in a binary classification problem are "is bubble" and "not bubble": "is bubble" represents the occurrence of market bubbles or financial collapse, and "not bubble" is the nonoccurrence. 

However, binary classification does not distinguish between market bubbles and financial collapse if a bubble period is identified. Financlal bubbles are preceded by a long meteoric rise in asset prices, and market collapses are also preceded by a step-down in the price of assets. 

We extend the binary classification to a multilabel classification. Here, two more classes are added: "is bubble asset creation" and "is bubble financial collapse". "is bubble asset creation" predicts bubble which results from the rampant creation of wealth, and "is bubble financial collapse" predicts bubbles resulting from financial crises and falling asset prices. Multilabel, and not multiclass classification, because an observation can be classified as "is bubble" and "is bubble asset creation" or "is bubble" and "is bubble financial collapse": An observation has to be classified as a bubble before it can be classified under "asset creation" or "financial collapse". Throughout the document, "bubble asset creation" refers to "bubble up", and "financial collapse" refers to "bubble down". Bubble up is determined by 

\begin{align}
\label{is_bubble_up}
	\text{is bubble up} = \begin{cases}
		1  &, \text{is bubble = 1} ~ \& ~ \frac{Y_k + \ldots + Y_{k+\tau}}{ \tau} > W\\
		0 &, \text{otherwise}
	\end{cases}
\end{align}
and bubble down is determined by
\begin{align}
\label{is_bubble_down}
	\text{is bubble down} = \begin{cases}
		1  &, \text{is bubble = 1} ~ \& ~ \frac{Y_k + \ldots + Y_{k+\tau}}{\tau} \le  W\\
		0 &, \text{otherwise}
	\end{cases}
\end{align}
The average of $\tau$ observations is compared with $W,$ where $W$ is a constant. 

%Likewise, bubble up and bubble down can be calculated using the current series value instead of an arbitrary constant $W$. Bubble up using this approach is
%
%\begin{align}
%	\text{is bubble up} = \begin{cases}
%		1  &, \text{is bubble = 1} ~ \& ~ Y_{k+\tau + 1} > \frac{Y_k + \ldots + Y_{k+\tau}}{ \tau} \\
%		0 &, \text{otherwise}
%	\end{cases}
%\end{align}
%and bubble down is 
%\begin{align}
%	\text{is bubble down} =\begin{cases}
%		1  &, \text{is bubble = 1} ~ \& ~ Y_{k+\tau + 1} \le \frac{Y_k + \ldots + Y_{k+\tau}}{ \tau} \\
%		0 &, \text{otherwise}
%	\end{cases}
%\end{align}

where $\tau$ is the window size. 

\subsection{Financial Sentiment Analysis with finbert-lc}
The second step of the three-stage analysis is to calculate polarity scores of financial texts. We use the fine-tuned finbert-lc model in \cite{atsiwo2024financial} to calculate polarity scores. Polarity score is a numerical score that indicates the sentiment of a piece of text. Polarity score takes values in the range $-1$ to $1$ inclusively. A polarity score near 1 signifies a positive sentiment, a score around $-1$ reflects a negative sentiment, and a score approximately 0 indicates a neutral sentiment.

Let Table \ref{bubble_prediction::three-stage-analysis::table:probability_distribution} be the probability distribution of a financial news title such that $p_1 + p_2 + p_3 = 1.$
\begin{table}[htbp]
	\centering
	\begin{tabular}{@{}lll@{}}
		\toprule
		Sentiment Class & Probability & Value \\ \midrule
		Negative        & $p_1$       & $-1$  \\
		Positive        & $p_2$       & $1$   \\
		Neutral         & $p_3$       & $0$   \\ \bottomrule
	\end{tabular}  
	\caption{Probability Distribution of Sentiment Classes for a Single Financial News Title.}
	\label{bubble_prediction::three-stage-analysis::table:probability_distribution}
\end{table}

The polarity score is the expected value under the probability distribution shown in Table \ref{bubble_prediction::three-stage-analysis::table:probability_distribution}, that is, 
\begin{equation}
	\begin{aligned}
		\text{polarity-score} &= (-1)p_1 + (1)p_2 + (0)p_3 \\
		&= p_2 - p_1
	\end{aligned}
\end{equation}
A polarity score of 1 implies that $p_1 = p_3 \approx 0$ and a polarity score of $-1$ implies thatn $p_2 = p_3 \approx 0.$ A polarity score of $0$ implies that $p_3 \approx 1$ and $p_1=p_2 \approx 0.$ The total polarity and average polarity scores for the financial news titles of $N$ are the sum and average of the individual polarity scores. The total polarity score is 
\begin{align}
	\text{total polarity-score} = \sum_{i=1}^N \text{polarity score}_i
\end{align}
and the average polarity score is 
\begin{align}
	\text{average polarity-score} = \frac{1}{N} \sum_{i=1}^N \text{polarity score}_i
\end{align}

\subsection{Ensemble Method for Robust Prediction}
The third-stage in the three step analysis is ensemble method for robust prediction. Replace ensemble models with any other machine learning model, and the three-stage analysis is still valid, but with a different predictive model. Ensemble models combine several weak learners to form a strong learner. 

An ensemble method integrates multiple weak classifiers to form a single robust classifier.  The weak classifiers are combined sequentially or independently. Combining sequentially or independently depends on the specific ensemble method. Some examples of ensemble approaches are Random Forest (RF), Extreme Gradient Boosting (XGBoost), and Adaptive Boosting (Adaboost). Random forest combines weak learners independently, and extreme gradient boosting and adaboost combine weak learners sequentially. Weak learners are trained sequentially with XGBoost, where each learner compensates for the weakness of the previous learners. Throughout this section, the implementation will be based on a decision tree as a weak learner. 

% A gentle introduction to the decision tree algorithm is provided. We provide details on the implementation of the random forest classifier. See Appendix \ref{appendix::three-stage-analysis} for implementation details on XGBoost and Adaboost. Details on some non-ensemble methods are also provided in the Appendix \ref{appendix::three-stage-analysis}.  

\subsection*{Boosting and Bagging}
Boosting combines multiple weak learners to create a strong learner. The key concepts paramount to the implementation of boosting are sequential learning, weighted data, and combining learners. 
With sequential learning, models are built sequentially where the current model corrects the errors of the previous model. Weights are assigned to the data, where misclassified observations receive high weights and correctly classified observations receive low weights. The high weights ensure that more misclassified examples are "seen" by the base learner on the next iteration. The base learners are combined into a weighted sum to form a strong learner. Boosting improves accuracy and reduces overfitting. 

\begin{figure}[h!]
	\centering
	\includegraphics[width=0.9\linewidth]{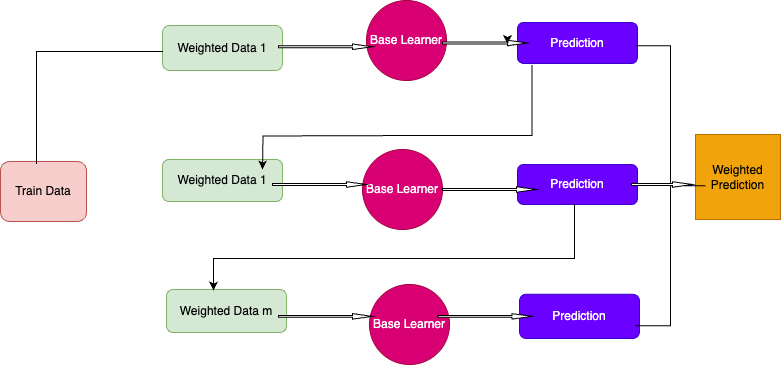}
	\caption{Boosting algorithm. Base learners are trained on weighted samples, where each base learner correct the errors of the previous learner. }
	\label{appendix::three-stage-analysis::boosting}
\end{figure}

Bagging (also known as boostrap aggregation), like boosting, combines base learners to form a strong learner. Bagging involves parallel learning, bootstrapping, and combining learners. Unlike boosting, base learners are fitted independently to bootstrap data. The independent base learners are combined to form an ensemble classifier. Bagging reduces variance and avoids overfitting.

\begin{figure}[h!]
	\centering
	\includegraphics[width=0.9\linewidth]{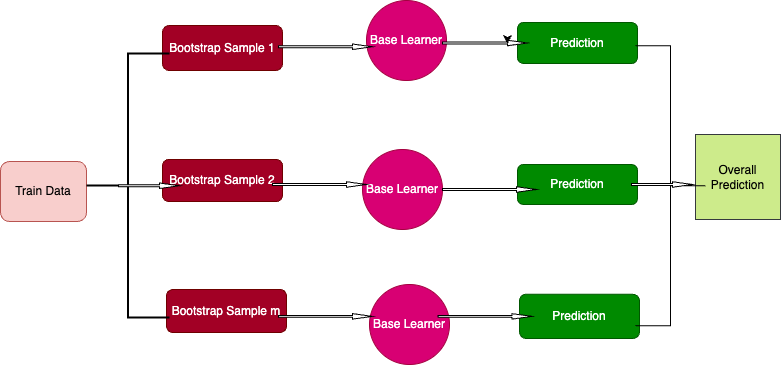}
	\caption{Bagging algorithm. Bootstrap samples are generated from the original sample and $m$ classifiers are trained independently.}
	\label{appendix::three-stage-analysis::boosting}
\end{figure}

\subsection*{Decision Tree}
\begin{figure}[h!]
	\centering
	\includegraphics[width=1\linewidth]{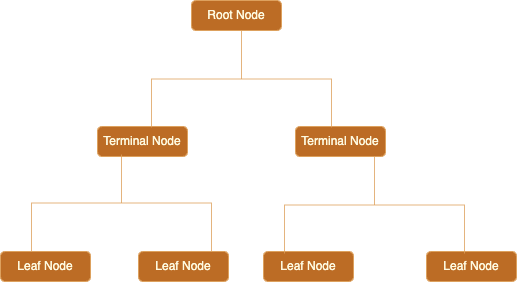}
	\caption{A decision tree hierarchy from the root node to the terminal node}
	\label{appendix::three-stage-analysis::decision-tree}
\end{figure}

A decision tree \cite{murthy1998automatic} is a supervised non-parametric learning algorithm that uses a recursive approach to group a target variable into a homogeneous group. A decision tree can be used for regression and classification tasks. The term Classification and Regression Trees (CART) in the literature refers to a decision tree for classification and regression tasks. A decision tree is shown in Figure \ref{appendix::three-stage-analysis::decision-tree}. The root node starts the recursion and the leaf node terminates the recursion if a split condition is satisfied. The terminal nodes divide the decision tree by the feature with the best score. 

The steps in fitting a decision tree to sample data are i) select the best feature to split, ii) split the dataset based on the feature with the best score, and iii) build tree recursively, and continue ii) until a leaf node is reached.

Let $D_m, ~\theta = (j, t_m)$ and $n_m$ represent the data, the candidate split and the sample size at node $m$, respectively. Here, $j$ is the feature and $t_m$ is the threshold. The threshold is determined from the sample points. 

The candidate split at each node is calculated with an impurity or loss function. The impunity is given as 
\begin{align}
	H(D_m) = \sum_k p_{mk}(1 - p_{mk})
\end{align}
and the entropy is calculated as 
\begin{align}
	H(D_m) = - \sum_k p_{mk}\log p_{mk}
\end{align}
where 
\begin{align}
	p_{mk} = \frac{1}{m} \sum_{y \in D_m} \textbf{I}(y = k)
\end{align} 

Given the candidate split, the data $D_m$ are divided into $D_{m}^L (\theta)$ and $D_{m}^R (\theta)$ subsets,  the left branch and the right branch, respectively. 
\begin{align}
	D_{m}^L (\theta) = \{ (x, y) | x_j \le t_m \}
\end{align}
for the left branch, and 
\begin{align}
	D_{m}^R (\theta) = { D_m \setminus D_m^L }
\end{align}
for the right branch. Given the left and right branches, the overall quality at node $m$ is calculated using 
\begin{align}
	\label{appendix::three-stage-analysis::decision-tree::quality}
	G(D_m, \theta) = \frac{n_m^L}{n_m} H(D_m^L, \theta) + \frac{n_m^R}{n_m} H(D_m^R, \theta)
\end{align}
with 
\begin{align}
	\theta^\star = \arg \min_{\theta} G(D_m, \theta)
\end{align}

Finally, recurse (repeat step ii) for subsets $H(D_m^L, \theta^\star)$ and $H(D_m^R, \theta^\star)$ until a maximum allowable depth is reached: $n_m < \text{min samples}$ or $n_m = 1.$

% The decision tree algorithm is given in Algorithm \ref{alg:decision_tree}. 

% The time complexity of training a balanced decision tree is $O(n_{samples}.n_{features} \log(n_{samples}))$. 

\section{Experimental Results}
\subsection{Data Sources and Processing}
We apply the three-stage analysis to predict bubbles with real-world financial time series. The financial time series used in this application is the S\&P 500 price-dividend (PD) ratio. The PD ratio of a stock is the ratio of the current price of a stock to the dividend paid to its investors over a period of time. A higher PD ratio implies that the stock is expensive compared to the amount of dividend paid. A lower PD ratio means that the stock is undervalued. Investors seeking dividend-paying stock monitor this metric closely.
The first step was to detect bubbles using the PSY test for bubble detection, which has the capacity to detect multiple bubble periods while taking into consideration multiplicity and heteroskedasticity. The PSY procedure outputs 1 (is bubble) if a bubble is detected and 0 (not bubble) if a bubble is not detected. The other two labels (which are is bubble up and is bubble down) were created with the formulas in Equations \ref{is_bubble_down} and \ref{is_bubble_up}.

The PSY method was used for monthly price-dividend ratio data ranging from January 1, 1940, to July 1, 2023, covering 1003 monthly observations. We used observations from January 1, 1940, to December 31, 1950, as training data for the PSY test. The actual bubble data start from January 1, 1951. The PSY test requires a minimum window size, which was determined using a rule in \cite{phillips2020real}. The identified bubble periods from the PSY test coincided, not surprisingly, with famous stock market crashes: Black Monday crash, Dot-com bubble, Covid-19 market crush, among others. 

\begin{figure}[h!]
	\centering
	\includegraphics[width=0.9\textwidth, height=0.4\textheight]{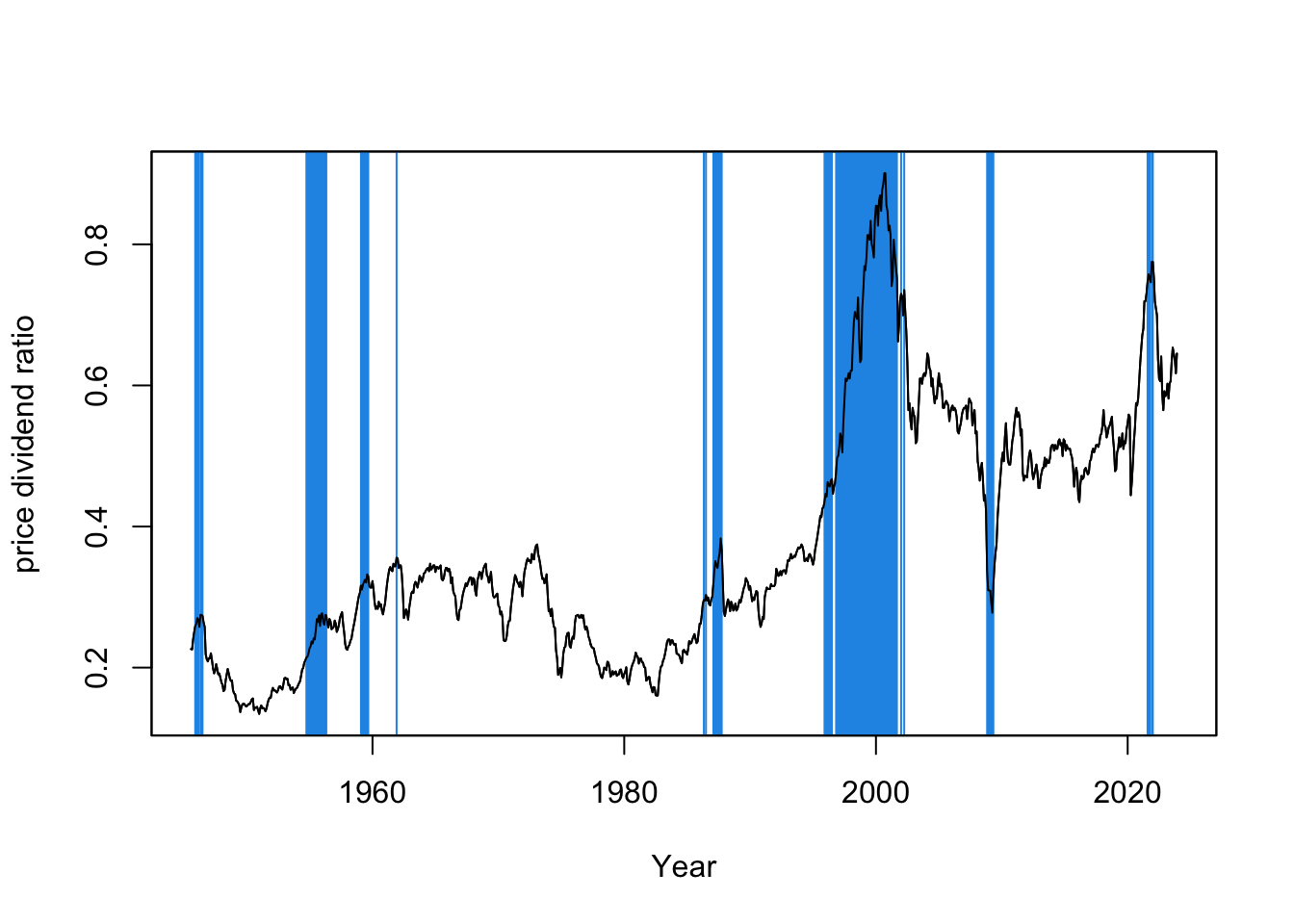}
	\caption{S\&P 500 price-dividend ratio with detected bubble periods drawn with vertical lines for monthly data}
\end{figure}

Short-term interest rate measures the cost of borrowing on financial instruments that mature in few years. The rates are determined by the central bank's monetary and economic policies. The three-month rate (3M T-Bill) was used as a measure of the short-term interest rate. They are usally daily averages based on the money market. Long-term interest rate (10Y T-Note), on the other hand, measures the cost of borrowing on financial instruments with longer maturities, practically from several years to decades. Long-term interest rates are common with mortages, loans, government, and private bonds. Interest rates can be either fixed or variable: fixed rates stay the same during its duration and offer a sense of security to its stakeholders; floating interest rates adjust to the current market conditions. 
Some factors that affect these interest rates are fiscal and monetary policies, inflation, economic growth policies, and global market conditions. Investors' buying and selling decisions are determined by these factors. 
Inflation is a measure of the percentage increase in a basket of goods and services within an economy during a particular period of time. It can be used as a tool to measure the strength of a currency. Inflation can be caused by the increase in the cost of production, shortages (demand exceeding supply), and others. It can be used as a tool to measure the purchasing power of a currency. An excessive inflation rate decreases the strength of a country's currency. 

The unemployment rate (U-Rate) is the percentage of the labor force that is unemployed and actively seeking for employment within a specified time. The labor force refers to those who are currently employed or are currently seeking employment within a specified time frame. The unemployment rate can be used as a tool for accessing the health of an economy.  Gross domestic product measures the total value of goods and services produced within the enclosure of a country in a given period of time. Let $C, I, G, E$ and $I$ be consumption, investment, government spending, exports, and imports, respectively. GDP can be expressed as 
\begin{align*}
	\text{GDP} = C + I + G + (X-M)
\end{align*}
where $X-M$ is the net export. 

Finally, the Balance of Payment (BOP) is a summary of all financial interactions with residents of one country with the rest of the world. These interactions occur through various accounts: current , capital and financial accounts.  The current account deals with the trade in of goods and services and transfers of income through the current account; the capital account keeps track of transactions related to the purchase and sale of assets (physical and electronic); and the financial account keeps track of financial transactions such as loans and investment.

It is difficult to access textual data from 2010 and below without scapping websites, making sentiment analysis on stock price movement challenging. 
The news data was collected from four different sources: Bloombery, Reuters, New York Times (NYT) and an API request from a third-party data vendor \cite{eodhd}.  Reuters and Bloombery data were compiled and used in \cite{BloombergReutersDataset2015}. The Reuters textual data have $109,110$ news titles from January 2010 to August 2016 and the Bloombery data has $450,341$ news contents. Financial news for Apple, Microsoft, Amazon, Meta, Tesla, JP Morgan Chase, Google, and Alphabet. These companies contribute over $50\%$ to S\&P 500 total earnings between 2015 and 2023. Finally, financial news was collected from NYT between 1960 and 2023 with the archive and search APIs. The search query 
\begin{verbatim}
	https:/api.nytimes.com/svc/search/v2/articlesearch.json?
	q=financials&api-key=yourkey
\end{verbatim} returns a json object of current financial news and the query 

\begin{verbatim}
	https:/api.nytimes.com/svc/archive/v1/2019/1.json?
	api-key=yourkey
\end{verbatim} returns a json object with all articles for Jan 2019.

The original data have $1007$ monthly data points, including 128 bubbles and 813 without bubble data. 66 monthly data points were used as training data for the PSY algorithm. The 128 bubbles have 65 and 63 bubble up and bubble down observations, respectively. The monthly data were converted into biweekly series by interpolating. The interpolation was applied on the PD ratio, and the PSY procedure was applied on the interpolated data. 

Long- and short-term interest rates are reported monthly. On the other hand, gross domestic product, unemployment rate, BOP, and inflation series are reported on a quarterly basis. These variables were converted to unformally to biweekly series. The bubble data and economic variables were interpolated with cubic spline interpolation because of its smooth behavior and the quality of the interpolation.

\subsection{Evaluation Metrics and Results}
We compare the performance of ensemble methods (XGBoost, Adaboost, Random Forest) with decision tree as a base learner, and non-ensemble methods like neural network, K nearest neighbor (KNN), naive bayes and logistic regression on different metrics. The results for 5-fold and 10-fold cross-validation with the f1 macro as scoring metric are reported in Tables \ref{three-stage-analysis::data_application::table::comparison_5foldcross valiaton} and \ref{three-stage-analysis::data_application::table::comparison_10foldcross valiaton}. 

\begin{table}[h!]
	\begin{tabular}{@{}llcccc@{}}
		\toprule
		Model Category                                                                              & Model               & \multicolumn{1}{l}{CV Score} & \multicolumn{1}{l}{F1 Score} & \multicolumn{1}{l}{Accuracy} & \multicolumn{1}{l}{CPU Minutes} \\ \midrule
		\multirow{4}{*}{\begin{tabular}[c]{@{}l@{}}Ensemble / Tree\\ Based \\ Methods\end{tabular}} & XGBoost             & 0.96                         & \textbf{0.97}                & \textbf{0.98}                & 63                              \\
		& AdaBoost            & 0.97                         & 0.96                         & 0.97                         & 2                               \\
		& Random Forest       & 0.97                         & 0.96                         & 0.98                         & 65                              \\
		& Decision Tree       & 0.96                         & 0.94                         & 0.95                         & 2                               \\ \midrule 
		\multirow{4}{*}{\begin{tabular}[c]{@{}l@{}}Non Ensemble \\ Methods\end{tabular}}            & K-Nearest Neighbor  & 0.91                         & \textbf{0.94}                & \textbf{0.97}                & 1                               \\
		& Logistic Regression & 0.24                         & 0.23                         & 0.75                         & 2                               \\
		& Naive Bayes         & 0.49                         & 0.48                         & 0.62                         & 1                               \\
		& Neural Network      & 0.92                         & 0.94                         & 0.93                         & 177                             \\ \midrule
	\end{tabular}
	\caption{Comparison of machine learning methods, categorized into ensemble / tree based methods (XGBoost, Random Forest, AdaBoost, Decision Tree) and non-ensemble methods (Neural Network, K-Nearest Neighbors, Logistic Regression, Naive Bayes), evaluated based on 5-fold cross-validation score, accuracy, and F1 score, and the time spent in parameter grid search.}
	\label{three-stage-analysis::data_application::table::comparison_5foldcross valiaton}
\end{table}
It is evident from Table \ref{three-stage-analysis::data_application::table::comparison_5foldcross valiaton} that KNN with K = 3 has the f1 score and accuracy, which requires 1 minute of CPU time. Neural network has the same f1 score as KNN with an accuracy of $93\%$ compared to $97\%$ for KNN. f1 macro provides a better measure of performance as the data is imbalanced. However, a neural network requires approximately 177 minutes CPU time, which is approximately 3 hours on a single CPU and 18 minutes on 10 CPU's.

For the tree-based methods, Extreme Gradient Boosting (XGBoost) outperformed the other tree based metric in terms of f1 score and accuracy. Random forest has the same accuracy value as XGBoost and takes approximately the same time to train. Ensemble methods have the same or higher f1 scores compared to non-ensemble methods.

\begin{table}[h!]
	\begin{tabular}{@{}llcccc@{}}
		\toprule
		Model Category                                                                              & Model               & \multicolumn{1}{l}{CV Score} & \multicolumn{1}{l}{F1 Score} & \multicolumn{1}{l}{Accuracy} & \multicolumn{1}{l}{CPU Minutes} \\ \midrule
		\multirow{4}{*}{\begin{tabular}[c]{@{}l@{}}Ensemble / Tree\\ Based \\ Methods\end{tabular}} & XGBoost             & 0.95                         & \textbf{0.97}                & \textbf{0.98}                & 125                             \\
		& AdaBoost            & 0.97                         & 0.96                         & 0.97                         & 4                               \\
		& Random Forest       & 0.97                         & 0.96                         & \textbf{0.98}                         & 134                             \\
		& Decision Tree       & 0.96                         & 0.93                         & 0.95                         & 3                               \\ \midrule 
		\multirow{4}{*}{\begin{tabular}[c]{@{}l@{}}Non Ensemble \\ Methods\end{tabular}}            & K-Nearest Neighbor  & 0.90                         & \textbf{0.94}                & \textbf{0.97}                & 1                               \\
		& Logistic Regression & 0.24                         & 0.23                         & 0.75                         & 3                               \\
		& Naive Bayes         & 0.49                         & 0.48                         & 0.62                         & 1                               \\
		& Neural Network      & 0.93                         & 0.95                         & 0.95                         & 422                             \\ \midrule
	\end{tabular}
	\caption{Comparison of machine learning methods, categorized into ensemble / tree based methods (XGBoost, Random Forest, AdaBoost, Decision Tree) and non-ensemble methods (Neural Network, K-Nearest Neighbors, Logistic Regression, Naive Bayes), evaluated based on 10-fold cross-validation score, accuracy, and F1 score, and the time spent in parameter grid search.}
	\label{three-stage-analysis::data_application::table::comparison_10foldcross valiaton}
\end{table}

The result of 10-fold cross-validation is summarized in Table \ref{three-stage-analysis::data_application::table::comparison_10foldcross valiaton} with different metrics compared. Like 5-fold cross-validation, XGBoost outperformed all models using f1 score as the scoring criteria and KNN outperforms other non-ensemble methods for both accuracy and f1 score. Unlike 5-fold cross-validation, KNN outperforms decision tree for both f1 score and accuracy. Neural network requires approximately 7 hours of CPU time, which is equivalent to 42 minutes of training on 10 CPUs and 420 minutes training on a single CPU.

The difference in comparing ensemble methods for five-fold and 10-fold cross-validation is subtle. Training XGBoost with 63 minutes CPU is within our compute resources, thus that is the ensemble model of choice.

\subsection{Feature Importance and Model Diagnostic, \textbf{RQ1}}
\begin{figure}
	\centering
	\includegraphics[width=1\linewidth]{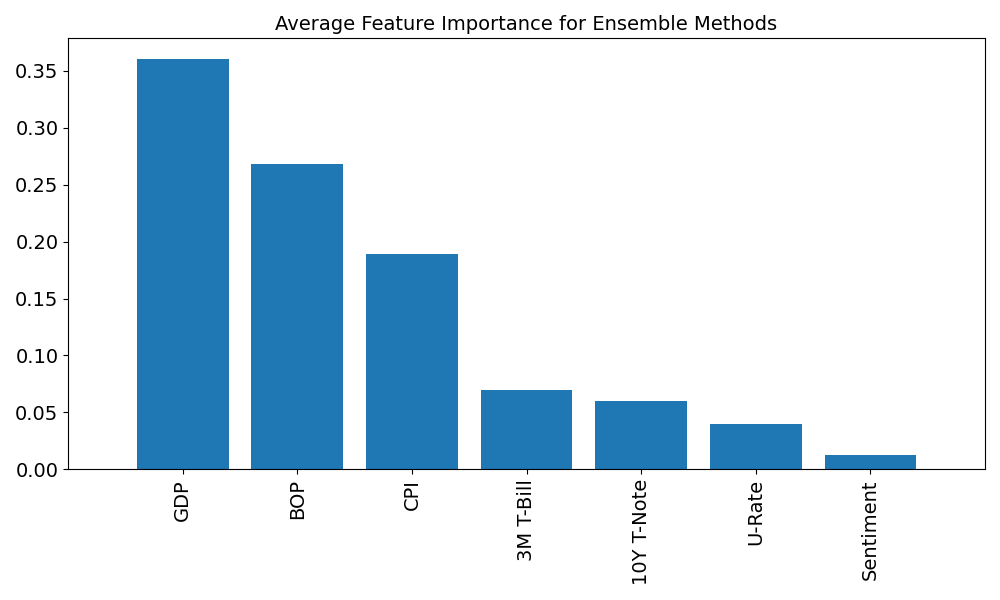}
	\caption{Feature Importance of the Average of Ensemble Methods}
	\label{three-stage-analysis::data_application::figure::comparison_feature_importance_extreme}
\end{figure}

\begin{figure}
	\centering
	\includegraphics[width=1\linewidth]{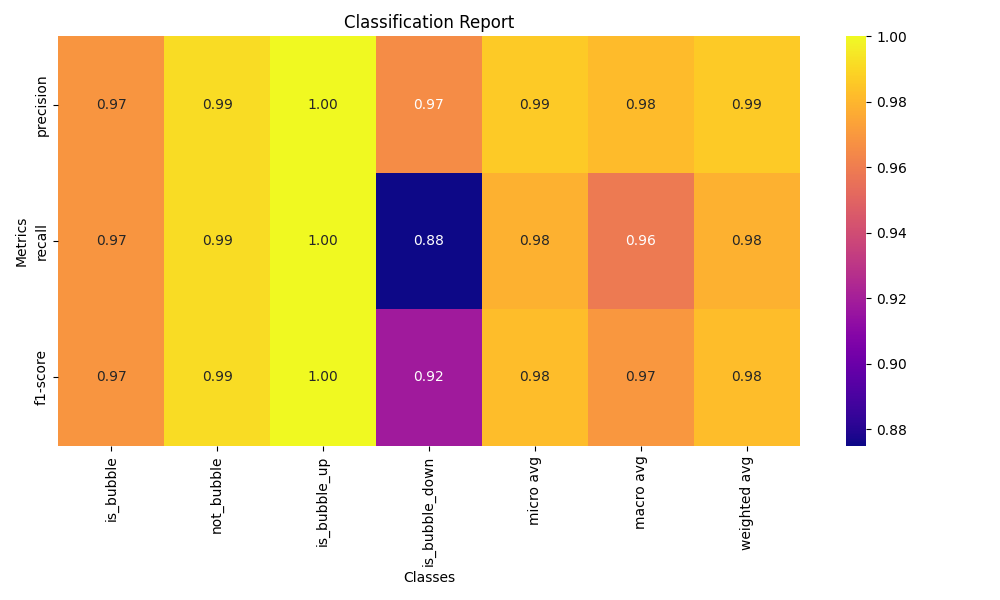}
	\caption{Heatmap of the Classification Report for Using XGBoost with a Threshold 0.5.}
	\label{three-stage-analysis::data_application::figure::class_metrics_comparison}
\end{figure}

Ensemble methods have features importance built into their implementation. Features that are used more often during the decision-making process (splitting) are considered more important than features that are used less. Feature importance is available for each group; however, we report the average feature importance across all labels for each model. The features are ranked according to their importance score, which is summarized in Table \ref{three-stage-analysis::data_application::table::comparison_feature_importance_extreme}. In bubble prediction (considering all classes), sentiment (calculated using new titles from financial news) is the least important feature. It ranks lowest for all ensemble models. GDP ranks first for all models except the Random Forest. GDP, CPI and BOP rank first to third, in some order, for all the ensemble methods, and 10Y T-Note, 3M T-Bill, and U-Rate rank fourth to sixth. 
\begin{table}[h!]
	\begin{tabular}{@{}llllllc@{}}
		\toprule
		Model                     & XGBoost    & Adaboost   & Random Forest & Decision Tree & Average    & \multicolumn{1}{l}{Rank} \\ \midrule
		\multirow{7}{*}{Features} & GDP        & GDP        & BOP           & GDP           & GDP        & 1                        \\
		& BOP        & CPI        & CPI           & BOP           & BOP        & 2                        \\
		& CPI        & BOP        & GDP           & CPI           & CPI        & 3                        \\
		& 10Y T-Note & 3M T-Bill  & 10Y T-Note    & 3M T-Bill     & 3M T-Bill  & 4                        \\
		& 3M T-Bill  & 10Y T-Note & 3M T-Bill     & U-Rate        & 10Y T-Note & 5                        \\
		& U-Rate     & U-Rate     & U-Rate        & 10Y T-Note    & U-Rate     & 6                        \\
		& Sentiment  & Sentiment  & Sentiment     & Sentiment     & Sentiment  & 7                        \\ \midrule
	\end{tabular}
	\caption{Feature importance scores and rankings across various ensemble methods.  Each method’s features and corresponding ranks provide insights into the relative influence and significance of each feature in the models’ predictions.}
	\label{three-stage-analysis::data_application::table::comparison_feature_importance_extreme}
\end{table}
The average column is calculated by averaging the important scores across the four models. The rank by averaging the columns is consistent with the rank of the individual models.  The top three pedictors of the occurrence or non-occurrence of a bubble are GDP, BOP, and CPI. It can be seen from Figure \ref{three-stage-analysis::data_application::figure::comparison_feature_importance_extreme} that GDP, BOP and CPI combine for more than $75\%$ of the importance scores. The combined metrics are reported in Tables \ref{three-stage-analysis::data_application::table::comparison_5foldcross valiaton} and \ref{three-stage-analysis::data_application::table::comparison_10foldcross valiaton}. Combined metrics (accuracy and f1 score) measure overall performance by averaging metrics from individual classes. Recall, precison, and f1 score for the individual classes ("is bubble", "not bubble", "is bubble up" and "is bubble down") are reported in Figure \ref{three-stage-analysis::data_application::figure::class_metrics_comparison}.

Of all the instances that the XGBoost model predicted as "is bubble", $97\%$ of the predictions are correct. When the model predicts the occurrence of a bubble, it is always almost correct. A high precision for this class implies a low false positive rate. Of the actual instances of "is bubble", XGBoost correctly identified $97\%$ of them, which implies a low false negative rate. The model strives for a good balance between precision and recall with an f1 score of $97\%$. A precision, recall and f1 score of $99\%$ is also impressive. Here, we are interested in the non-occurrence of a bubble. If the model predicts that a bubble will not occur, it is true $99\%$ of the time, and $99\%$ of the "not bubble" instances are correctly predicted by the model. If a bubble occurs, the likelihood of it being on the higher side is what we call "is bubble up". All predicted and actual instances are correctly identified by the model, resulting in an f1 score of $100\%$. The reverse of "is bubble up" is "is bubble down", the likelihood of a bubble ending up on the lower side taking into consideration the threshold. We examine the impact of different thresholds in Section \ref{three-stage-analysis::data_application::section::where-model-fails}, varying threshold and where the model fails.

\begin{figure}[h!]
	\centering
	\includegraphics[width=1.15\textwidth, height=0.6\textheight]{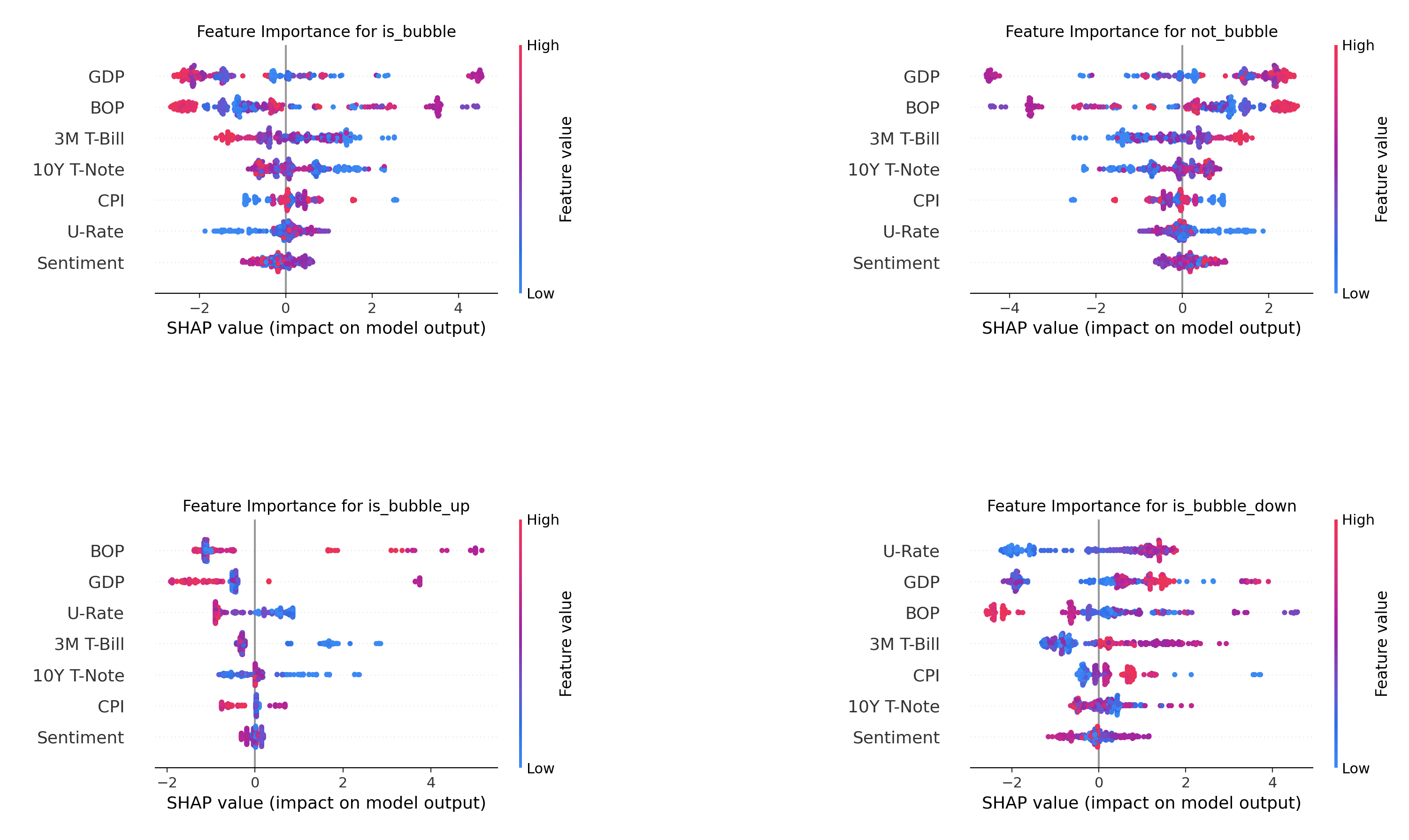}
	\caption{SHAP summary plots for feature importance in multilabel classification on test data. 
	}
	\label{three-stage-analysis::data_application::figure::shap_class_metrics_comparison}
\end{figure}

The SHAP plot for all four classes is shown in Figure \ref{three-stage-analysis::data_application::figure::shap_class_metrics_comparison}. The x-axis and the y-axis represent the shap values, representing the impact on the prediction and the features, ordered by their importance (from top to bottom), respectively. The color gradient explains how individual features contribute to the final prediction of instances. GDP and BOP are essential features to predict the test instances. Like feature importance, sentiment calculated using financial news does not contribute to the prediction.

\subsection{Varying Threshold and Where the Model Fails, \textbf{RQ3}}
\label{three-stage-analysis::data_application::section::where-model-fails}

We examine the impact of varying the threshold on precision, recall, and f1 score in each class. The confusion matrix measures where the model fails and instances that need careful consideration.

\begin{table}[h!]
	\centering
	\begin{tabular}{llcccccccc}
		\toprule
		bubble label & metric & RM & C=0.2 & C=0.3 & C=0.4 & C=0.5 & C=0.6 & C=0.7 & C=0.8 \\
		\midrule
		is bubble & precision & 0.97 & 0.95 & 0.94 & 0.91 & 0.97 & 0.97 & 0.97 & 0.95 \\
		& recall & 0.92 & 0.97 & 0.94 & 0.91 & 0.97 & 0.98 & 0.91 & 0.94 \\
		& f1 score & 0.94 & 0.96 & 0.94 & 0.91 & 0.97 & 0.98 & 0.94 & 0.95 \\ \midrule
		& support & 65 & 65 & 65  & 65 & 65 & 65 & 65 & 65 \\ \midrule 
		not bubble & precision & 0.98 & 0.99 & 0.98 & 0.97 & 0.99 & 1.00 & 0.98 & 0.98 \\
		& recall & 0.99 & 0.99 & 0.98 & 0.97 & 0.99 & 0.99 & 0.99 & 0.99 \\
		& f1 score & 0.99 & 0.99 & 0.98 & 0.97 & 0.99 & 0.99 & 0.98 & 0.99 \\ \midrule 
		& support & 238 & 238 & 238 & 238 & 238 & 238 & 238 & 238 \\ \midrule 
		is bubble up & precision & 0.79 & 0.95 & 0.95 & 0.95 & 1.00 & 1.00 & 0.95 & 1.00 \\
		& recall & 0.79 & 0.95 & 0.96 & 0.93 & 1.00 & 1.00 & 0.95 & 1.00 \\
		& f1 score & 0.79 & 0.95 & 0.96 & 0.94 & 1.00 & 1.00 & 0.95 & 1.00 \\ \midrule 
		& support & 39 & 63 & 56 & 42 & 33 & 29 & 21 & 8 \\ \midrule 
		is bubble down & precision & 0.60 & 1.00 & 0.75 & 0.91 & 0.97 & 0.94 & 0.95 & 0.95 \\
		& recall & 0.46 & 0.50 & 0.67 & 0.91 & 0.88 & 0.94 & 0.93 & 0.93 \\
		& f1-score & 0.52 & 0.67 & 0.71 & 0.91 & 0.92 & 0.94 & 0.94 & 0.94 \\ \midrule
		& support & 26 & 2 & 9 & 23 & 32 & 36 & 44 & 57 \\ \midrule
		\bottomrule
	\end{tabular}
	\caption{Measuring the performance of XGBoost for "is bubble, not bubble, is bubble up" and "is bubble down" using the Rolling Mean (RM) Approach with a window width of three and by Varying the Threshold. precsion, recall and f1 score are recorded for each threshold and label.}
	\label{three-stage-analysis::data_application::table::varying_threshold}
\end{table}

There are 65 bubble data points and 238 nonbubble data points (Table \ref{three-stage-analysis::data_application::table::varying_threshold}). The number of examples for "is bubble up" and "is bubble down" depends on the threshold. 63 instances are on the high side and 2 instances are on the low side for a decision boundary (threshold) of 0.2. On the other hand, 8 examples are on the high side and 57 examples are on the low side for a decision boundary of 0.8. A high decision boundary reduces the number of data points on the high side, and a low decision boundary reduces the number of data points on the low side if there is a bubble. 

The decision boundaries of 0.2 and 0.3 have a recall (precision) of 0.50 (0.67) and 0.67 (0.71), respectively. Increasing the decision boundary decreases the false positive and true positive rates. A boundary of at least 0.4 suffices for splitting "is bubble" into "is bubble up" and "is bubble down". The choice of $C$ depends on the specific application and the cost of incorrect predictions for each class.

\begin{figure}[h!]
	\centering
	\includegraphics[width=1\linewidth]{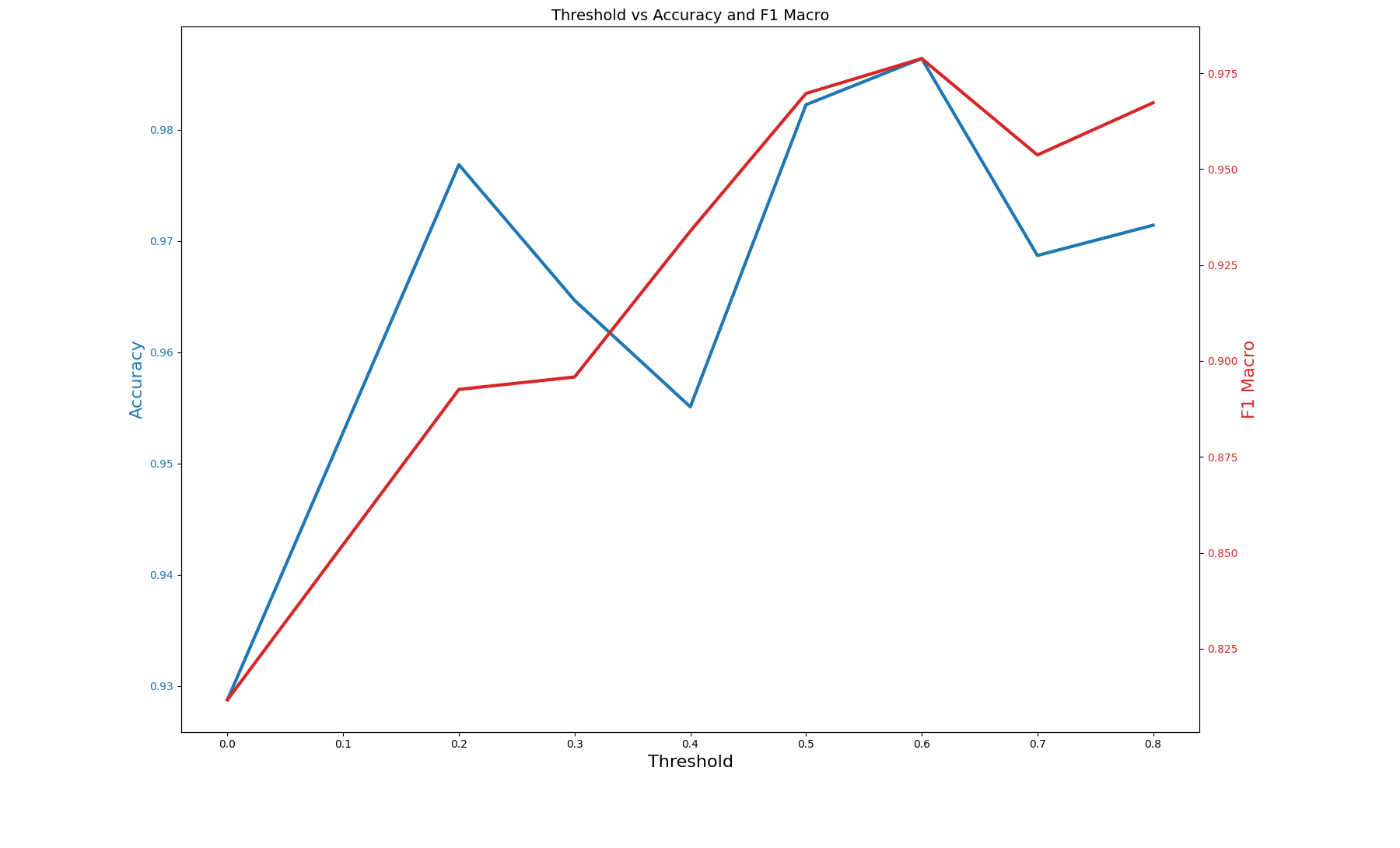}
	\caption{Micro and macro averaging compared for different threshold. }
	\label{three-stage-analysis::data_application::figure::threshold_accuracy_f1score_comparison}
\end{figure}

Rolling mean approaches compare the mean of the previous $w$ observations with the current observation, where $w$ is the width of the window. An observation is classified as "is bubble up" if the current observation is greater than the rolling mean. It is classified as "is bubble down" otherwise. The precision, recall and f1 score for the said labels with this approach is low. The dynamics with the bubble direction when this is a bubble is not correctly captured by the rolling mean decision boundary technique. Figure \ref{three-stage-analysis::data_application::figure::threshold_accuracy_f1score_comparison} compares micro-averaging (accuracy) and macro-averaging for different thresholds.

Next, we examine where the model fails by examining the multilabel confusion matrix for the default threshold of 0.5, which is summarized in Table \ref{three-stage-analysis::data_application::table::multilabel_confusion_matrices}.

\begin{table}[H]
	\centering
	\begin{tabular}{cc}
		% First Row
		\begin{tabular}{cc|c|c|}
			\cline{3-4}
			& is bubble & \multicolumn{2}{c|}{\textbf{Predicted}} \\
			\cline{3-4}
			& & 0 & 1 \\
			\hline
			\multicolumn{1}{|c|}{\multirow{2}{*}{\textbf{Actual}}} & 0 & 236 & 2 \\
			\multicolumn{1}{|c|}{} & 1 & 2 & 63 \\
			\hline
		\end{tabular}
		&
		\begin{tabular}{cc|c|c|}
			\cline{3-4}
			& not bubble & \multicolumn{2}{c|}{\textbf{Predicted}} \\
			\cline{3-4}
			& & 0 & 1 \\
			\hline
			\multicolumn{1}{|c|}{\multirow{2}{*}{\textbf{Actual}}} & 0 & 63 & 2\\
			\multicolumn{1}{|c|}{} & 1 & 2 & 236 \\
			\hline
		\end{tabular}
		\\ \\
		% Second Row
		\begin{tabular}{cc|c|c|}
			\cline{3-4}
			& is bubble up& \multicolumn{2}{c|}{\textbf{Predicted}} \\
			\cline{3-4}
			& & 0 & 1 \\
			\hline
			\multicolumn{1}{|c|}{\multirow{2}{*}{\textbf{Actual}}} & 0 & 270 & 0 \\
			\multicolumn{1}{|c|}{} & 1 & 0 & 33 \\
			\hline
		\end{tabular}
		&
		\begin{tabular}{cc|c|c|}
			\cline{3-4}
			& is bubble down & \multicolumn{2}{c|}{\textbf{Predicted}} \\
			\cline{3-4}
			& & 0 & 1 \\
			\hline
			\multicolumn{1}{|c|}{\multirow{2}{*}{\textbf{Actual}}} & 0 & 270 & 1\\
			\multicolumn{1}{|c|}{} & 1 & 4 & 28 \\
			\hline
		\end{tabular}
	\end{tabular}
	\caption{Confusion matrices for multilabel classification for the default threshold.}
	\label{three-stage-analysis::data_application::table::multilabel_confusion_matrices}
\end{table}

Two non bubble observations and two bubble observations are incorrectly classified as bubble and non bubble respectively. All instances are correctly classified for the label "is bubble up". For the label "is bubble down", 4 positive instances are incorrectly predicted as negative, and a single negative example is incorrectly predicted as positive. XGBoost does a good job of predicting the rare classes: "is bubble down, is bubble up and is bubble".
\section{Conclusion and Further Work}

In this paper, we developed a three-step machine learning approach to predict bubbles in the S\&P 500 market by integrating financial news sentiment with key macroeconomic indicators. The proposed framework combined a right-tailed unit root test for bubble detection, sentiment extraction using the model developed in \cite{atsiwo2024financial}, and ensemble-based prediction using multiple supervised learning algorithms.

The experimental results showed that while the inclusion of financial news sentiment provided only a marginal improvement in predictive performance, macroeconomic variables remained the dominant drivers of bubble formation. Among the ensemble methods, XGBoost achieved the highest F1-score, demonstrating a superior balance between precision and recall in identifying bubble periods. For non-ensemble models, k-Nearest Neighbors (k-NN) yielded the best performance, although its computational cost and sensitivity to noise made it less scalable for larger datasets.

The ranking of variable importance indicated that GDP growth, balance of payments (BOP), and consumer price index (CPI) were the most influential predictors of bubble dynamics, followed by 3-month Treasury Bill rates, 10-year Treasury Note yields, unemployment rate, and news sentiment. These findings reinforce the key role of macroeconomic fundamentals in bubble prediction, while sentiment acts as a complementary signal that captures short-term market reactions.

Overall, the proposed three-step approach demonstrated high predictive power and robustness, making it a practical tool for policymakers, regulators, and investors seeking early warning signals of market instability. Future research could explore more advanced sentiment modeling or cross-market extensions to assess whether global financial news or international economic linkages enhance the accuracy of bubble prediction models. Size-based variables used in \cite{atsiwo2024capital} can also be included in the set of predictors.

%\clearpage
%\bibliographystyle{plain}
%\bibliographystyle{alpha}
%\bibliography{references}

\end{document}